\documentclass[iop,apj]{emulateapj}

\usepackage{graphicx}
\usepackage{hyperref}
\usepackage{amssymb,amsmath}

\def\lea{\mathrel{<\kern-1.0em\lower0.9ex\hbox{$\sim$}}}
\def\gea{\mathrel{>\kern-1.0em\lower0.9ex\hbox{$\sim$}}}
\newcommand{\lta}{{\>\rlap{\raise2pt\hbox{$<$}}\lower3pt\hbox{$\sim$}\>}}
\newcommand{\gta}{{\>\rlap{\raise2pt\hbox{$>$}}\lower3pt\hbox{$\sim$}\>}}

\begin{document}
\title{Multiwavelength study of the northeastern outskirts of the extended TeV source HESS J1809$-$193}

\author{Blagoy Rangelov\altaffilmark{1}, Bettina Posselt\altaffilmark{2}, Oleg Kargaltsev\altaffilmark{1}, George G. Pavlov\altaffilmark{2}, Jeremy Hare\altaffilmark{1} and Igor Volkov\altaffilmark{1,3}}
\altaffiltext{1}{Department of Physics, The George Washington University, 725 21st St, NW, Washington, DC 20052}
\altaffiltext{2}{Pennsylvania State University, 525 Davey Lab., University Park, PA 16802}
\altaffiltext{3}{University of Maryland, College Park, MD 20742 }
\email{rangelov13@gwu.edu}

\slugcomment{The Astrophysical Journal, in press}
\shorttitle{The Northeastern Outskirts of HESS~J1809$-$193}
\shortauthors{Rangelov et al. 2014}

\begin{abstract}

HESS~J1809$-$193 is an extended TeV $\gamma$-ray source in the Galactic Plane. Multiwavelength observations of the HESS~J1809$-$193 field reveal a complex picture. We present results from three \emph{CXO} and two \emph{Suzaku} observations of a region in the northeastern outskirts of HESS J1809--193, where enhanced TeV emission has been reported. Our analysis also includes GeV $\gamma$-ray and radio data. One of the X-ray sources in the field is the X-ray binary XTE~J1810--189, for which we present the outburst history from multiple observatories and confirm that XTE~J1810--189 is a strongly variable type I X-ray burster, which can hardly produce TeV emission. We investigate whether there is any connection between the possible TeV extension of HESS~J1809$-$193 and the sources seen at lower energies. We find that another X-ray binary candidate, \emph{Suzaku}~J1811--1900, and a radio supernova remnant, SNR~G11.4$-$0.1, can hardly be responsible for the putative TeV emission. Our multiwavelength classification of fainter X-ray point sources also does not produce a plausible candidate. We conclude that the northeast extension of HESS~J1809$-$193 , if confirmed by deeper observations, can be considered as a dark accelerator -- a TeV source without visible counterpart at lower energies.

\end{abstract}


\section{Introduction}

During the past decade observations with TeV $\gamma$-ray observatories, such as High Energy Stereoscopic System (HESS), revealed a large number of very-high energy (VHE) sources in the Galactic plane \citep{Aharonian2005A}. Sources, firmly identified as pulsar-wind nebulae (PWNe), shell-type supernova remnants (SNRs), and microquasar-type high-mass X-ray binaries (HMXBs) account for about a half of the total number ($\sim90$) of Galactic VHE sources \citep{Kargaltsev2013}. There is a large number of unidentified VHE sources ($\sim20$), for which multiwavelength (MW; from radio to TeV) observations provide hints of a counterpart (such as an SNR interacting with a molecular cloud, or a star-forming region). Most of these associations are still uncertain because at least some of these sources still could  be powered by offset pulsars whose PWNe are faint in X-rays. Among these is a group of  6-7 ``dark'' sources which do not show plausible counterparts at any other wavelengths.

\begin{figure*}
\includegraphics[scale=0.325]{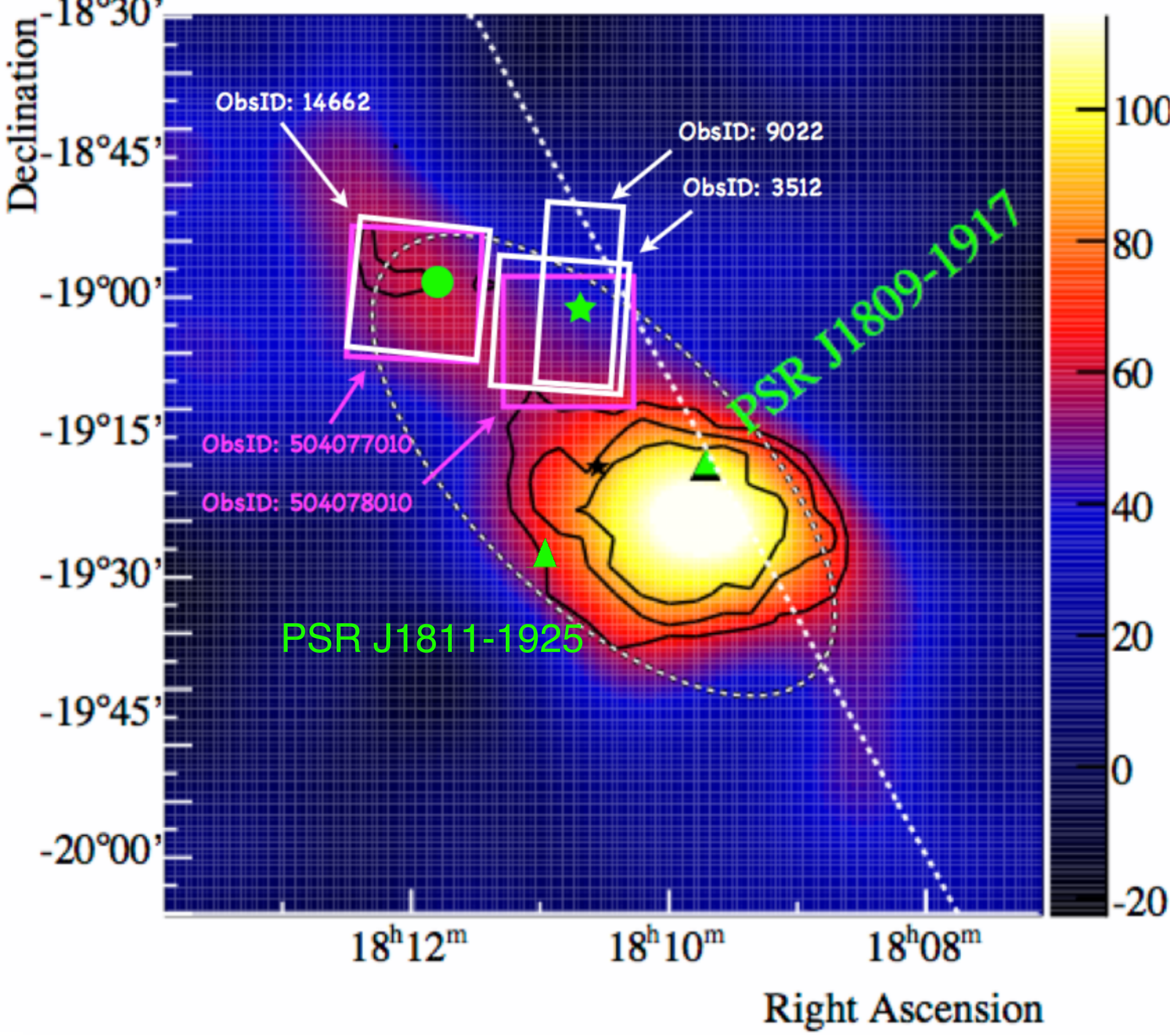}~
\includegraphics[scale=0.32,trim=0 -90 0 0]{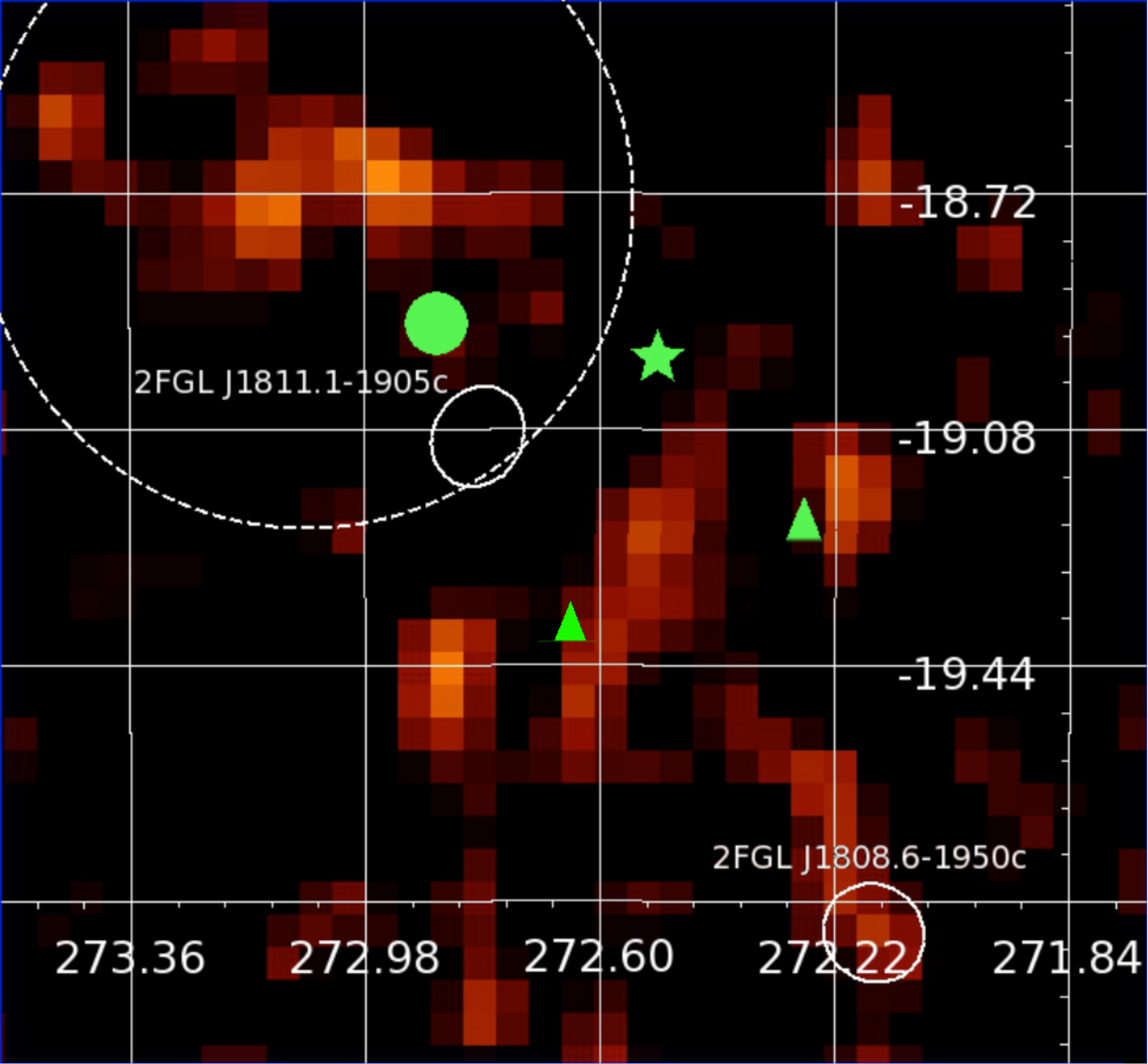}
\caption{\emph{Left}: An image of HESS~J1809$-$193 in $1-10$~TeV (adopted from \citealt{Aharonian2007}), smoothed with a Gaussian of width $6\farcm6$. The color scale is set such that the blue/red transition occurs at approximately the 3~$\sigma$ significance level. The black contours are the 4, 5 and 6~$\sigma$ significance contours. The position of the pulsars PSR~J1809--1917 and PSR J1811$-$1925 are marked with green triangles, the two X-ray sources, \emph{Suzaku}~J1811-1900 and XTE~J1810--189, with green circle and star, respectively. The Galactic plane is shown as a white dotted line. The best-fit position for the $\gamma$-ray source is marked with a black star and the fit ellipse with a dashed line. The purple and white rectangles show the \emph{Suzaku} XIS and \emph{CXO} pointings, respectively. \emph{Right}: \emph{Fermi} test Statistic (TS) map of the HESS region in the 15--300~GeV range. \emph{Suzaku}~J1811-1900, XTE~J1810--189 and PSR~J1809--1917 are denoted as in the left panel. The \emph{Fermi} counts extraction region ($r=0.5^\circ$) is shown with white dashed circle, while the white ellipses represent 2FGL J1811.1-1905c and 2FGL J1808.6-1950c.; see Section~2.3.}
\end{figure*}

HESS~J1809$-$193 was observed as part of the systematic survey of the inner Galaxy  \citep{Aharonian2005A,Aharonian2006A}. As there was a marginally significant VHE $\gamma$-ray signal ($2\sigma$), further observations of HESS~J1809$-$193 were undertaken (with total live time of $\sim25$ hours), which resulted in a detection significance of $6.8\sigma$ for VHE $\gamma$-ray emission within $13'$ of the location of PSR J1809-1917 \citep{Aharonian2007}. The HESS image (shown in Figure 1) suggested a fainter extension northeast (NE) of the main source, which could be a separate source.  A year later,  \citet{Renaud}  presented preliminary results from a 41-hour HESS exposure of the same source. In Figure 1 from Renauld et al. (2008) the NE TeV extension appears to be more fragmented, with multiple blobs, which could be explained by multiple faint sources but their individual significances would be very low.

There are several possible sources of energetic particles known in this region, including two young pulsars.  Most of the HESS~J1809$-$193 TeV emission is likely produced by the PWN of the 51~kyr-old PSR~J1809$-$1917 via inverse Compton scattering (ICS; \citealt{Aharonian2007}). While the central region of HESS~J1809$-$193 has been investigated reasonably well \citep{Kargaltsev2007,Anada2010,Komin2008}, no comprehensive investigation has been done on the NE part despite the available MW coverage. Another energetic pulsar, J1811$-$1925, is more offset from the center of HESS~J1809$-$193 and it is also more distant than PSR~J1809$-$1917; $\sim5$~kpc compared to 3.5~kpc (ATNF catalog; \citealt{2005AJ....129.1993M}). Moreover, PSR J1811$-$1925 is located at the center of  SNR G11.2--0.3 whose size is much smaller than the pulsar's offset from the center of HESS~J1809$-$193. Therefore, the PWN of J1811$-$1925 cannot account for the TeV emission from the entire HESS~J1809$-$193  (see also \citealt{2008MNRAS.384L..29D}). PSR J1811$-$1925 is coincident with one of the TeV blobs seen by \citet{Renaud}, and its PWN could contribute some of the TeV emission from HESS~J1809$-$193; however, it is unlikely to be responsible for the $\gamma$-ray emission from the NE region of the HESS source (see Figure~1, left panel).

In this paper we focus on  the multiwavelength picture of the  region NE of  HESS~J1809$-$193, and investigate the nature of various sources seen in this region at lower energies. We also discuss whether any of them  could be sources of TeV emission.

We present the results from five X-ray observations $-$ three taken with the \emph{Chandra X-ray Observatory} (\emph{CXO}) and two with \emph{Suzaku} (see Figure~2). We analyze the brightest sources discovered in the \emph{CXO} and \emph{Suzaku} fields. One of these X-ray sources is the known low-mass X-ray binary (LMXB) candidate XTE~J1810$-$189, for which we show the outburst history from multiple observatories. The other one is a new X-ray binary (XRB) candidate \emph{Suzaku}~J1811--1900. We also provide MW classification for other fainter X-ray sources detected in the \emph{CXO} ACIS observations.

This paper is organized as follows. Section~2 summarizes the X-ray observations, and Section~3 presents the analysis of the X-ray data. Section~4 discusses the implications of these results for the production of $\gamma$-rays, and in Section~5 we summarize our main conclusions.

\section{Observations and Data Reduction}

\subsection{CXO}

We use three sets of archival \emph{CXO} observations of HESS~J1809$-$193 (Table~1). The data were taken with the ACIS-I instrument on board \emph{CXO} (ObsIDs 3512 and 14662) in ``very faint'' timed exposure mode, and with the HRC-S instrument in ``timing'' mode (ObsID 9022). We processed the data using the \emph{CXO} Interactive Analysis of Observations (CIAO\footnote{\url{http://cxc.harvard.edu/ciao/index.html}}) software (version 4.6) and \emph{CXO} Calibration Data Base (CALDB) version 4.5.9, and restricted the data to the energy range 0.5--8~keV. We use CIAOÕs Mexican-hat wavelet source detection routine \emph{wavdetect} \citep{Freeman2002} to detect X-ray sources and measure their coordinates in the \emph{CXO} images (listed in Table~2; Figure~2). CIAO's task \emph{srcflux} was used to extract net counts and model-independent source fluxes.

\begin{deluxetable}{lcccc}
\tablecaption{List of observations}
%\tabletypesize{\scriptsize}
\tablewidth{0pt}
\tablehead{
\colhead{Obs.} & \colhead{Date} & \colhead{ObsId} & \colhead{PI} & \colhead{Exp\tablenotemark{a}}}
\startdata
\emph{CXO} & 2003-10-18 & 3512 & Garmire & 20 \\
\emph{CXO} & 2008-04-23 & 9022 & Chakrabarty & 1 \\
\emph{CXO} & 2013-05-17 & 14662 & Posselt & 55 \\
\emph{Suzaku} & 2009-09-09 & 504077010 & Kargaltsev & 52 \\
\emph{Suzaku} & 2009-09-10 & 504078010 & Kargaltsev & 52 \\
\emph{Swift} & 2008-03-17 & 00031167001 & \nodata & 2 \\
\emph{Swift} & 2008-03-18 & 00031167002 & \nodata & 2 \\
\emph{Swift} & 2008-03-21 & 00031167003 & \nodata & 2 \\
\emph{Swift} & 2008-03-22 & 00031167004 & \nodata & 2 \\
\emph{Swift} & 2008-03-23 & 00031167005 & \nodata & 2 \\
\emph{Swift} & 2008-03-24 & 00031167006 & \nodata & 1 \\
\emph{Swift} & 2008-03-25 & 00306737000 & \nodata & 1 \\
\emph{Swift} & 2011-06-19 & 00455640000 & \nodata & 4 
\enddata
\tablenotetext{a}{Exposure in units of ks.}
\end{deluxetable}

\begin{deluxetable*}{lccrrrc}
\tablecaption{X-ray sources detected in the two \emph{CXO}/ACIS-I fields shown in Figure~2}
\tablewidth{0pt}
\tablehead{
\colhead{\#} & \colhead{RA} & \colhead{Dec} & \colhead{$F$\tablenotemark{a}} & \colhead{Net Counts} & \colhead{HR\tablenotemark{b}} & \colhead{Class\tablenotemark{c} (Probability)}}
\startdata
 1 & 273.04064 & --19.04081 & $15\pm1$ & $228\pm15$ & 0.91  & LMXB (71\%) \\
 2 & 273.01545 & --19.01186 & $2.0\pm0.5$ & $41\pm6$ & 0.69  & ?\\
 3 & 272.90919 & --19.01146 & $4.3\pm0.9$ & $74\pm9$  & 0.98  & YSO (89\%)\\
 4 & 273.01503 & --19.00379 & $2.3\pm0.4$ & $107\pm10$ & --0.43 & ?\\
 5 & 272.97179 & --19.00031 & $0.99\pm0.09$ & $41\pm6$  & -0.97 & ?\\
 6 & 273.01190 & --18.91413 & $9.6\pm1.0$ & $171\pm13$ & 0.57  & ?\\
 7 & 272.89484 & --18.96038 & $2.0\pm0.4$ & $81\pm9$  & --0.53 & STAR (99\%)\\
 8 & 272.90928 & --18.95289 & $2.2\pm0.7$ & $31\pm6$  & 0.96  & ? \\
 9 & 272.90288 & --19.06230 & $2.2\pm0.6$ & $50\pm8$  & 0.64  & ? \\
10 & 273.11526 & --18.93897 & $3.8\pm0.6$ & $131\pm12$ & --0.68 & STAR (99\%)\\
11 & 272.83877 & --18.92205 & $23\pm5$ & $76\pm10$  & 0.44  & ? \\
12 & 272.82446 & --18.97315 & $8.6\pm2.0$ & $44\pm7$  & 0.69  & ?\\
13 & 272.58428 & --19.08521 & $2.6\pm0.6$ & $43\pm7$  & --0.79 & STAR (99\%)\\
14 & 272.77570 & --19.09171 & $3.7\pm0.9$ & $40\pm6$  & 0.47  & ?\\
15 & 272.65398 & --19.05966 & $1.4\pm0.4$ & $36\pm6$  & --0.95 & ?\\
16 & 272.73350 & --19.08788 & $1.5\pm0.5$ & $31\pm5$  & --0.66 & STAR (99\%)
\enddata
\tablenotetext{a}{Model-independent X-ray fluxes in the $0.2-7$~keV range in units of $10^{-14}$ erg~s$^{-1}$~cm$^{-2}$.}
\tablenotetext{b}{Hardness ratio calculated as $(H-S)/(H+S)$, where $S$ and $H$ are the number of counts in the 0.2--2~keV and 2--7~keV bands, respectively.}
\tablenotetext{c}{Classification  and probability according to the automative classification algorithm.}
\end{deluxetable*}

\subsection{Suzaku}

The archival data from two \emph{Suzaku} observations of HESS~J1809$-$193 (see Table~1) were processed with FTOOLS' task XSELECT in the package HEASOFT\footnote{\url{http://heasarc.nasa.gov/lheasoft/}} version 6.13. We extract both PIN and GSO spectra from the HXD detector using the tasks \emph{hxdpinxbpi} and \emph{hxdgsoxbpi} respectively. We use appropriate PIN and GSO background HXD NXD files available in the archive\footnote{\url{ftp://legacy.gsfc.nasa.gov/suzaku/data/background/}}. No signal above the background is detected in the HXD detector in both observations.

\subsection{Fermi}

We use all archival \emph{Fermi} LAT data acquired between 2008 August 06 and 2014 June 10. The data were analyzed with the Fermi Science Tools following the standard procedures\footnote{\url{http://fermi.gsfc.nasa.gov/ssc/data/analysis}}.

\begin{deluxetable}{rcccccc}
\tablecaption{Outburst history of XTE~J1810-189}
\tablewidth{0pt}
\tablehead{
\colhead{Date} & \colhead{n$_{H}$\tablenotemark{a}} & \colhead{$\Gamma$\tablenotemark{b}} & \colhead{$F_{\rm{obs}}$\tablenotemark{c}} & \colhead{$F_{\rm{PL}}$\tablenotemark{d}} & \colhead{Exp.\tablenotemark{e}} & \colhead{Obs.}}%\\
\startdata
2003-10-18 & 4.0 & 1.7 & $<$8.6e-5 & $<$2e-4 & 20 & \emph{CXO} \\
2008-03-10 &1.0 &1.9&2.5&\nodata&\nodata & \emph{RXTE} \\
2008-03-17 &3.9 &1.7&2.9&~6.4&2.0 & \emph{Swift} \\
2008-03-18 & 4.0 &1.9&4.3&11&1.3 & \emph{Swift} \\
2008-03-21 & 5.0 &2.1 &3.1&11&1.8 & \emph{Swift} \\
2008-03-22 & 4.3 &1.7&3.4&~7.7&1.7 & \emph{Swift} \\
2008-03-23 & 4.6&1.9&4.3&13&1.8 & \emph{Swift} \\
2008-03-24 & 3.4&1.4&8.1&14&2.0 & \emph{Swift} \\
2008-03-25 & 4.0&1.5&4.5&~8.7&1.3 & \emph{Swift} \\
2008-03-26 & 1.0&1.9&2.5e2&\nodata&\nodata & \emph{RXTE} \\
2008-04-23 & 3.8&1.6&1.7&10&43 & \emph{CXO} \\
2009-09-11 & 2.0&1.6&5.8e-2&0.1&1 & \emph{Suzaku} \\
2011-06-19 & 4.9&2.6&6.5&45.5&3.8 & \emph{Swift} \\
2013-01-05 &\nodata&\nodata&9.6&\nodata&\nodata & \emph{MAXI} 
\enddata
\tablenotetext{a}{Hydrogen column density in units of $10^{22}$~cm$^{-2}$.}
\tablenotetext{b}{Power law photon index.}
\tablenotetext{c}{Observed X-ray flux in units of $10^{-10}$~erg~s$^{-1}$~cm$^{2}$ in 0.5$-$8~keV.}
\tablenotetext{d}{Unabsorbed X-ray flux in units of $10^{-10}$~erg~s$^{-1}$~cm$^{2}$ in 0.5$-$8~keV.}
\tablenotetext{e}{Exposure in units of ks.}
\end{deluxetable}

\subsection{Swift}

We use eight archival \emph{Swift} XRT observations (see Table~1) of XTE~J1810--189. The data were taken with the XRT instrument in PC mode. We processed the data using the online \emph{Swift}-XRT tool\footnote{\url{http://www.swift.ac.uk/user\_objects/}} \citep{Evans2007}.

\begin{figure*}
\begin{center}
\includegraphics[scale=0.34]{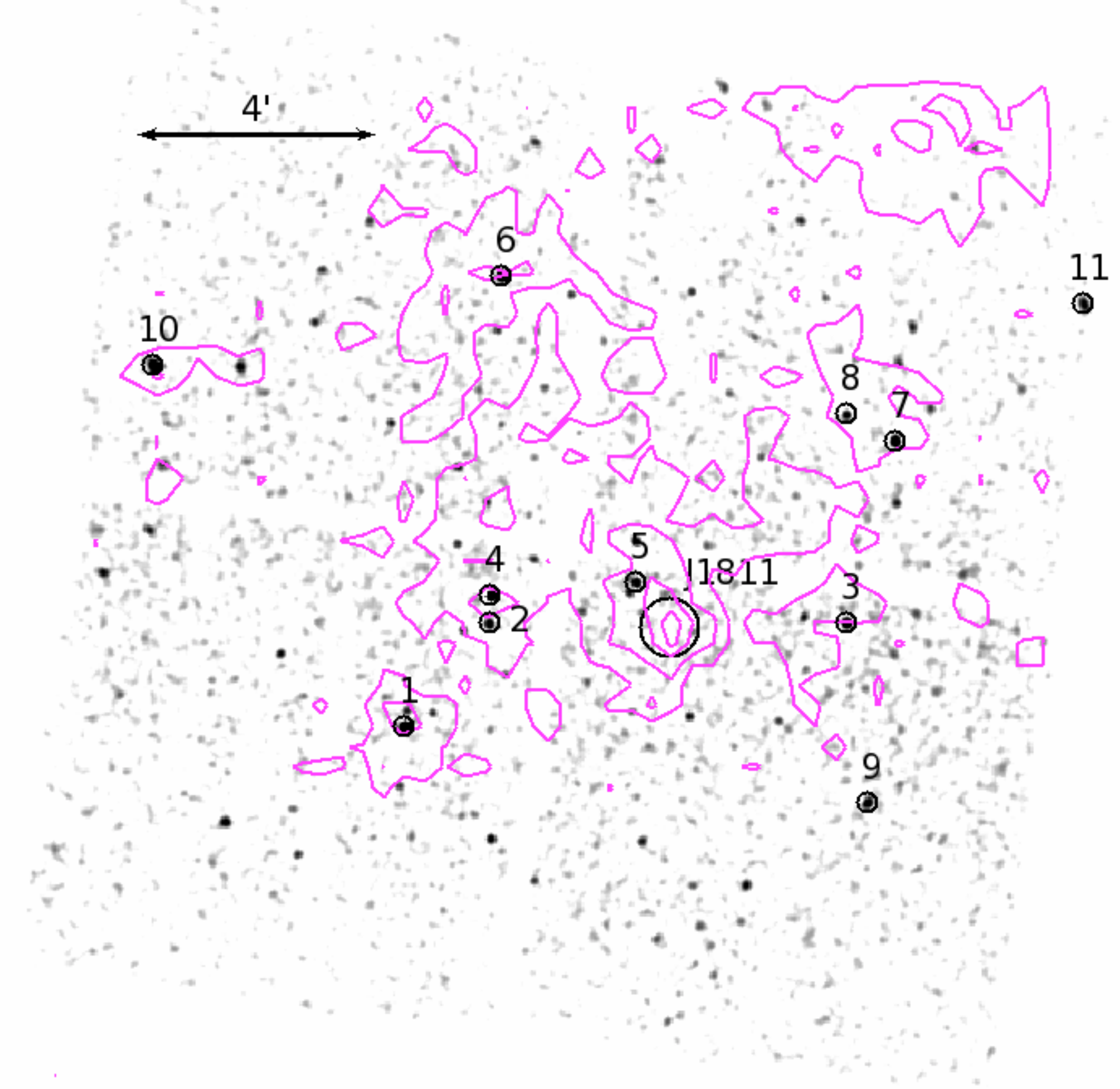}
\includegraphics[scale=0.48]{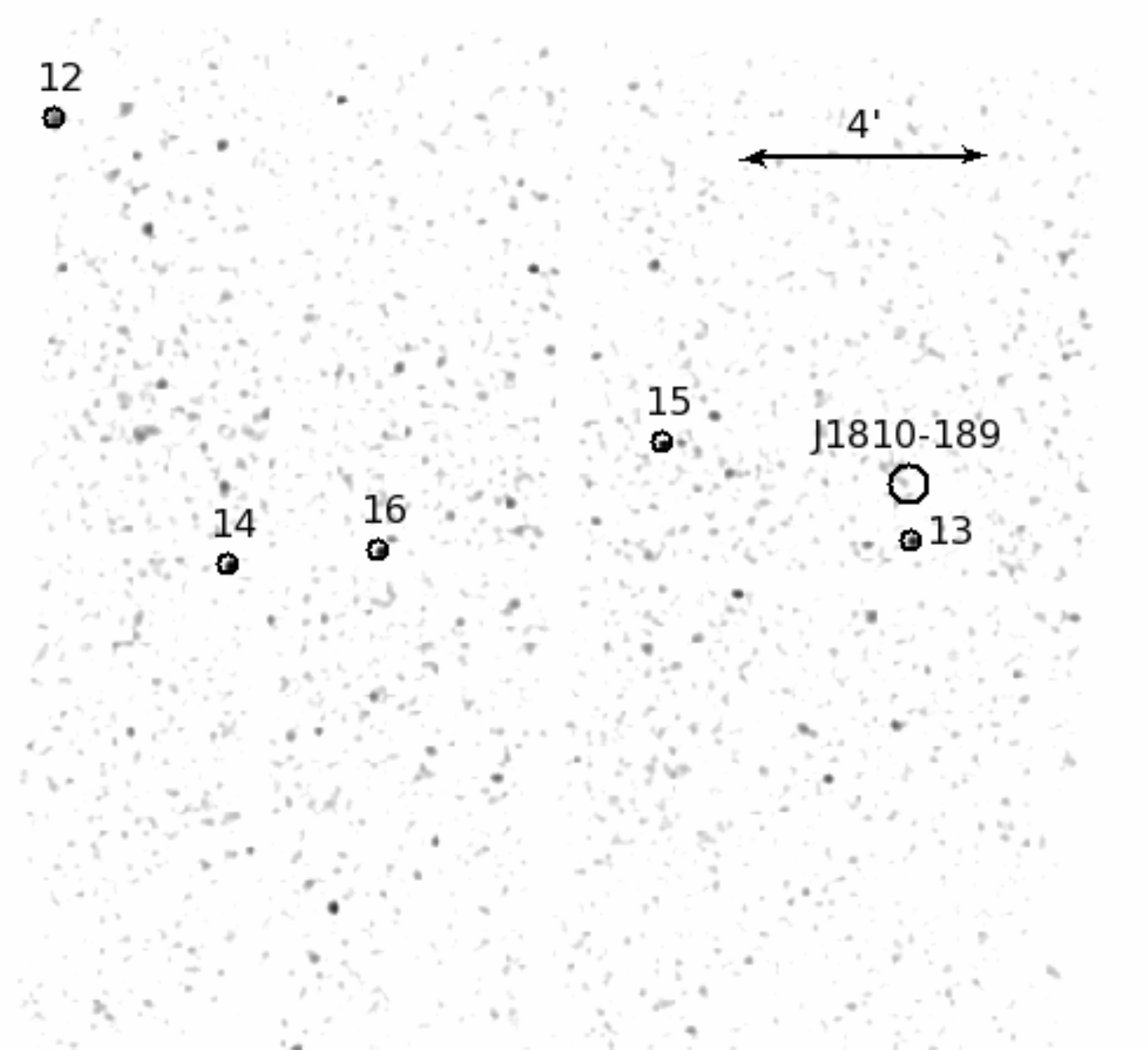}
\caption{\emph{CXO}/ACIS-I inverted scale images (ObsId 14662 (\emph{Left}) and ObsId 3512 (\emph{Right})) with 16 X-ray sources detected (Table~1). The positions of \emph{Suzaku}~J1811 and XTE~J1810--189 are also shown on top of the ACIS-I images, but neither of the two is detected. The magenta contours on the left panel trace the apparent extended emission of \emph{Suzaku}~J1811.}
\end{center}
\end{figure*}

\begin{figure}
\includegraphics[scale=0.59]{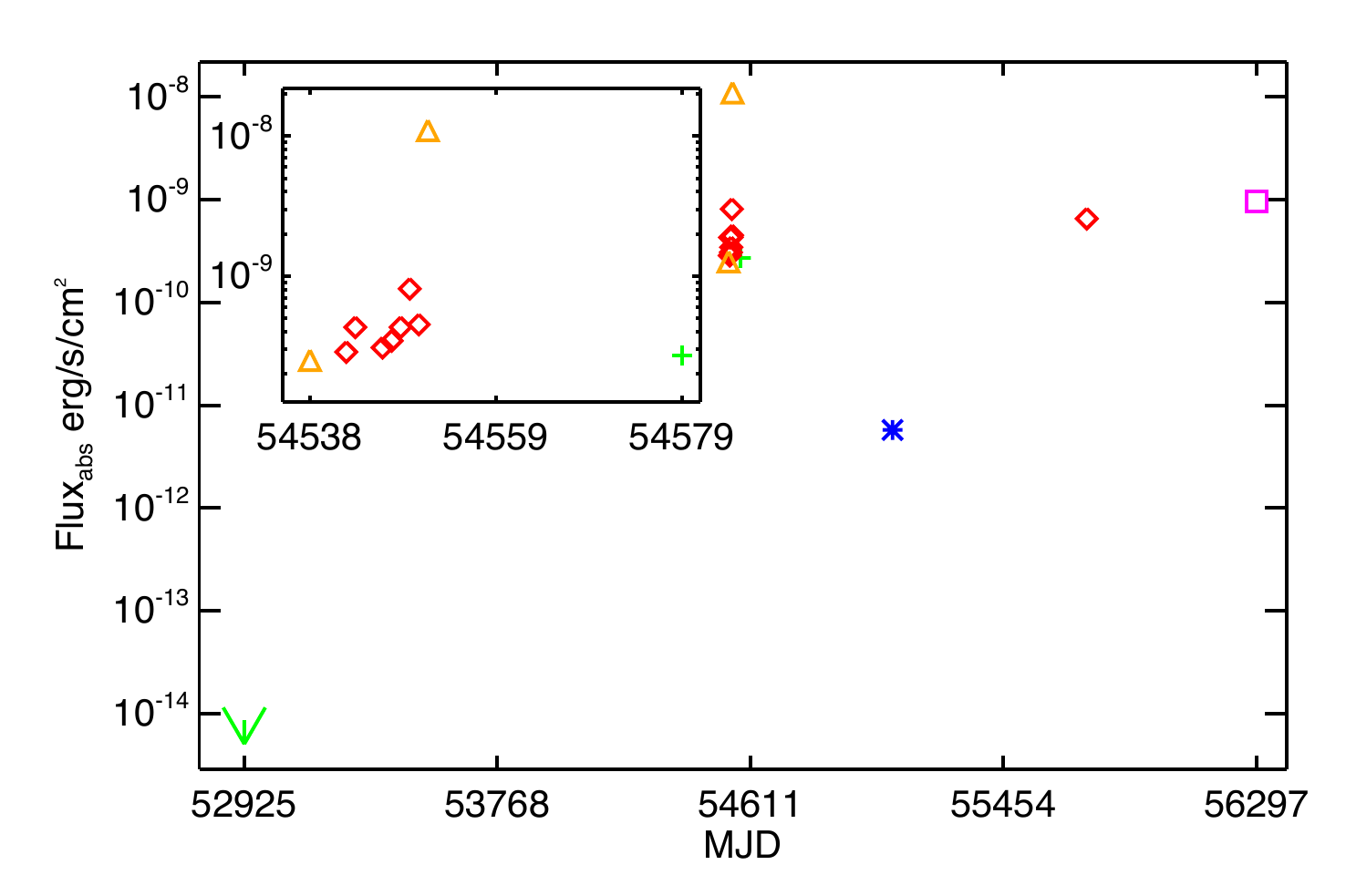}
\caption{Outburst history of XTE~J1810-189. Data from \emph{Swift} are shown with red diamonds, \emph{CXO} with green crosses (the arrow shows the upper limit), \emph{Suzaku} with blue asterisk, \emph{RXTE} \citep{Markwardt_Swank2008} with orange triangles, and \emph{MAXI} \citep{Negoro2013} with magenta square (see text for details). The inset shows the time during which the Type I X-ray burst occurred. The peak detected with \emph{RXTE} is clearly seen.}
\end{figure}

\section{Results}

We have detected 16 X-ray sources (with $\ge30$ net counts and detection significance $\ge10$) in two ACIS-I fields (Table~2). The X-ray spectra and responses were extracted with standard CIAO software. The fits were performed using XSPEC 12.8 in the $0.5-8$~keV energy range. For each source, we fitted an absorbed PL model.

\subsection{XTE~J1810$-$189}

We extracted source and background spectra for XTE~J1810--189 for both \emph{Suzaku} instruments XIS and HXD; however, no signal above the background was detected for the latter. The XIS spectrum is fit with an absorbed PL with photon index $\Gamma=1.6$ and $n_{H,22} = 2$ (where $n_{H,22}$ is the absorbing hydrogen column density in units of $10^{22}$~cm$^{-2}$). This corresponds to an unabsorbed X-ray flux of $5.8\times10^{-12}$~erg~cm$^{-2}$~s$^{-1}$ in the $0.5-10$~keV band.

We did not detect any X-ray sources in the 2003 October 18 ACIS-I image within $2\arcsec$ (\emph{CXO}'s positional uncertainty) of XTE~J1810$-$189. We determined an upper limit on the absorbed flux of $8.6\times10^{-15}$~erg~cm$^{-2}$~s$^{-1}$ using an absorbed PL model (assuming $n_{H,22}=4$ and $\Gamma=1.7$, average for all other observation).

The count rate of XTE~J1810--189 in the HRC-S observation was 1.86(4) counts~s$^{-1}$ \citep{Chakrabarty2008}. Assuming an absorbed PL X-ray spectrum with $\Gamma=2.0$ and  $n_{H,22}=3.8$ (as measured with 3--30 keV \emph{RXTE} PCA archival data from 2008 April 10 by \citealt{Torres2008}), \citet{Chakrabarty2008} estimated an absorbed (unabsorbed) X-ray flux of $2.7\times10^{-10}$ ($1\times10^{-9}$) erg~s$^{-1}$~cm$^{-2}$ in the $0.1-10$ keV band.

\citet{Markwardt_Swank2008} detected the variable source XTE~J1810$-$189 using \emph{RXTE}/PCA pointed observation on 2008 March 10. The authors modeled the spectrum with absorbed PL ($n_{H,22}= 1$ and $\Gamma=1.9$) and reported a 6.4~keV iron line, although a contamination by diffuse Galactic ridge emission could not be excluded. The X-ray flux, uncorrected for diffuse contamination, was $2.5\times10^{-10}$~erg~cm$^{-2}$~s$^{-1}$ in the 2--10~keV range.

An X-ray burst from XTE~J1810--189 was detected with \emph{RXTE}/PCA in a pointed observation on 2008 March 26 at 12:47 UT \citep{Markwardt2008}. A cooling trend of the thermal spectrum suggests a Type~I thermonuclear burst from a neutron star (NS). \citet{Markwardt2008} reported an unabsorbed peak X-ray flux of $\sim2.5\times10^{-8}$~erg~cm$^{-2}$~s$^{-1}$, which was used to obtain an upper limit of 11.5~kpc on the distance (assuming a standard Eddington peak luminosity of $3.8\times10^{38}$~erg~s$^{-1}$). 

XTE~J1810--189 experienced another outburst on 2013 January 5, with a flux of $~\sim40$~mCrab ($\sim9.6\times10^{-10}$~erg~cm$^{-2}$~s$^{-1}$) in the $4-10$~keV band, detected with MAXI/GSC \citep{Negoro2013}. The \emph{Swift}/BAT light curve, however, suggests that the outburst started on 2012 December 10. \citet{Negoro2013} reported that the source was in the hard state on 2013 January 5. The light curve is shown of Figure~3 (values listed in Table~3).

\citet{Torres2008} acquired a 900~s $K_S$-band image of the XTE~J1810--189 field on 2008 March 18 using the PANIC camera attached to the 6.5~m \emph{Magellan Baade} telescope. The seeing was $\sim0\farcs5$. The PSF-fitting photometry revealed that the brightest object (the NIR counterpart of XTE~J1810--189 proposed by the authors) within the \emph{CXO} error circle has declined in brightness from $K_S=17.3\pm0.2$ to $K_S=18.0\pm0.1$ (2008 June 23; \citealt{Torres2008_2}).

We used archival $V,H_{\alpha},R,I$ optical data from Cerro Tololo Inter-American Observatory (CTIO) 4-m \emph{Blanco} telescope taken on 2004 May 21. While the optical counterpart of XTE~J1810--189 was not detected in the observations, we estimate upper limits of $V>22.6$, $R>22.5$, $H_\alpha>19.6$, and $I>20.52$. However, this is not restrictive in terms of star spectral type because the non-detection can be attributed  entirely to the very large ISM absorption (A$_{\rm V}\ge11.5$; \citealt{2009MNRAS.395.1640R}).

\subsection{Suzaku X-ray Source}

In the 2009 \emph{Suzaku} image (ObsId 504077010) we discovered a ``compact'' X-ray source \emph{Suzaku} J1811--1900 (hereafter J1811) that appears to be marginally extended. It is apparently surrounded by large-scale extended emission (the respective contours are shown in Figure~2). Around 1000 (background subtracted) counts were collected from the compact \emph{Suzaku} source in the 52~ks exposure in a circular aperture with radius $80\arcsec$ at energies 0.5--8~keV.

We searched for periodicity of the compact source in the light curve. We have analyzed 514 photon arrival times from 3 XIS detectors extracted from a $r=30''$ region centered at R.A.=18:11:51, Decl.=--19:00:54 (which is the best-fit centroid of  the source). We corrected the arrival times to the solar system barycenter using \texttt{aebarycen} task. The arrival times were recorded with the resolution of 8\,s and spanned the interval of 99.94~ks. We searched for periodic signal using the Digital Fourier Transform and the $Z^{2}_n$ tests \citep{Buccheri1983}. No periodic signal with significance $>1.9\sigma$ was found  in the 0.00002--0.0625~Hz  range (for 6246 independent trials) we searched\footnote{There is a strong periodic signal at 96 min which is the {\sl Suzaku} orbital period.}. The maximum found $Z^{2}_1=23$ implies that the upper limits on $Z^{2}_1$ are 46, 49 and 60 at the 95\%, 99\% and 99.9\% confidence levels, respectively \citep{1975ApJS...29..285G}. These correspond to 40\%, 44\% and 48\% upper limits on the observed pulsed fraction (see, e.g., \citealt{1999ApJ...511L..45P}. The limits on the intrinsic pulsed fraction are a factor 1.6 larger. Therefore, the obtained upper limits on the pulsed fraction are not very restrictive.

\begin{figure}
\includegraphics[height=85mm, angle=-90]{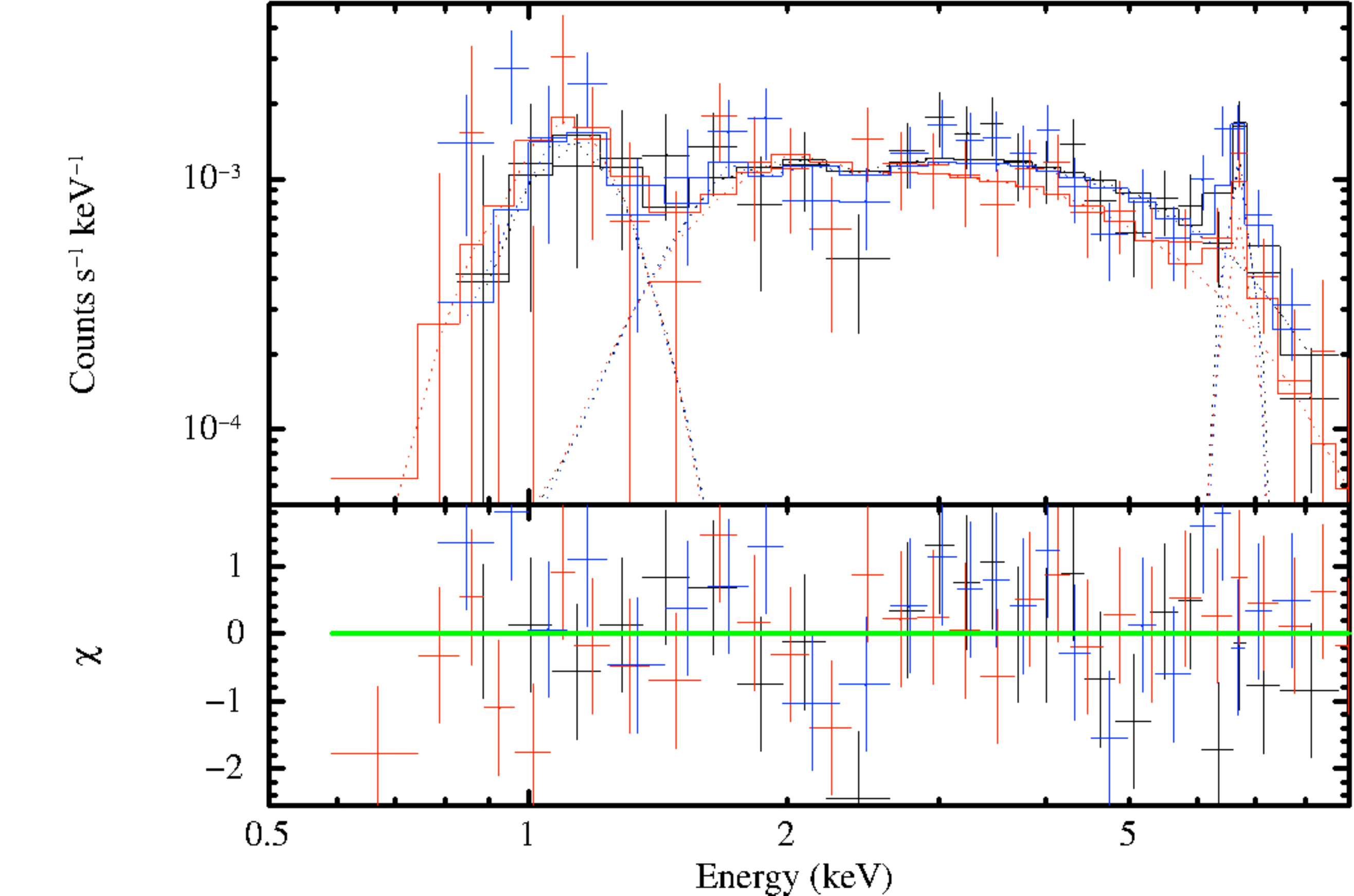}
\caption{X-ray spectrum of the \emph{Suzaku} J1811 source and the XSPEC fit with a model consisting of an absorbed combination of three components $-$ a PL, a blackbody, and a gaussian emission line (model components shown as dotted lines). The black, red, and blue points correspond to the XI0, XI1 and XI3 detector respectively.\label{Suzakuspec}}
\end{figure}

We fitted the spectrum of the compact source with an absorbed PL + blackbody (BB) + emission line model  (\texttt{gaussian} in XSPEC). For the photoelectric absorption in the interstellar medium, we used \texttt{tbabs} with the solar abundance table from \citet{Wilms2000} and the photoelectric cross-section table from \citet{Balu1992}. The best-fit (reduced\,$\chi^2=0.85$ for 71 dof) model parameters and their 90\% confidence levels are: $n_{H,22}=4.1^{+1.8}_{-1.4}$, $\Gamma=1.7^{+0.5}_{-0.4}$, $kT=60\pm 20$~eV, and a gaussian emission line at energy $6.70\pm0.07$~keV with a width of $\sigma=0.16^{+0.11}_{-0.09}$~keV, likely a Fe K$_\alpha$ line. The normalizations have large uncertainties, in particular, the 90\% confidence level of the BB normalization includes a range from $N_{BB}= R^2_{BB} D^{-2}_{10}= 3\times 10^5 D^{-2}_{10}$\,km$^2$ to $3\times 10^{10} D^{-2}_{10}$\,km$^2$, where $R_{BB}$ is the effective radius of an emitting equivalent sphere, and $D_{10}$ is the distance in units of 10\,kpc. Due to this large uncertainty, the size of the emission area is poorly constrained. The gaussian emission line contains ($4 \pm 2) \times 10^{-6}$ photons cm$^{-2}$\,s$^{-1}$, the PL normalization is $N_{PL}=8^{+10}_{-4} \times 10^{-5}$ photons\,keV$^{-1}$\,cm$^{-2}$\,s$^{-1}$ at 1 keV. The X-ray spectrum and the fit are shown in Figure~\ref{Suzakuspec}. The observed absorbed flux of J1811 is $3.4^{+0.2}_{-0.4} \times 10^{-13}$\,erg\,cm$^{-2}$\,s$^{-1}$ in the energy range 1--10\,keV.

The total Galactic H~I column density in this direction is $n_{HI}=1.8 \pm 0.5 \times 10^{22}$\,cm$^{-2}$ \citep{DickeyLockman1990}, a value slightly smaller than our $n_{H,22}$ which takes also molecular hydrogen into account. We note that \citet{2006ApJ...639..929B} found the X-ray $n_{H}$ values to be a factor of 2--3 greater than the 21~cm H~I column densities (for high Galactic column densities $\gtrsim10^{21}$~cm$^{-2}$).

We investigated standard surveys for coverage of J1811, e.g., at radio (VLA, SUMSS), optical (NOMAD, DSS), Galactic H$_{\alpha}$ (SuperCOSMOS), NIR (2MASS), IR (\emph{Spitzer},\emph{WISE}), hard X-ray frequencies (\emph{Integral}). There are no obvious correlations between emission in these bands and the \emph{Suzaku} emission around J1811. There are, however, many NIR point sources within the positional error circle ($\sim20\arcsec$) of \emph{Suzaku} $-$ around 30 cataloged 2MASS sources plus at least as many fainter objects visible in the 2MASS images. Several $H_\alpha$ point sources are within the error circle of the compact \emph{Suzaku}~J1811 source, too. There is no known X-ray binary within $30\arcsec$ of the Suzaku position in VizieR catalogs, e.g., \citet{Ritter2003}. A \emph{Spitzer} point-like source is situated $20''$ from \emph{Suzaku}~J1811. The source is bright ($K=6.7$) in the NIR (2MASS) and IR (\emph{Spitzer}), but it is not detected in the optical band. Comparison to isochrones reveals significant reddening, which cannot be explained solely by the Galactic extinction. This source is not coincident with any of the \emph{CXO} sources. In principle, this source -- or one of the fainter NIR sources within the $20''$ radius -- could be the counterpart of \emph{Suzaku}~J1811.

The \emph{Suzaku} source seems to be located within the NE extension of HESS~J1809$-$193. In standard astronomical databases (e.g., SIMBAD, the ATNF pulsar catalog, the \emph{Integral} catalog) there is neither a known pulsar nor a known galaxy (cluster) at the \emph{Suzaku} position. The position was covered by \emph{ASCA} at an off-axis angle of $\sim 14\arcmin$ in a 12~ks GIS observation. No obvious source is seen at the position of the compact source, indicating variability (around 60 counts are expected based on \emph{Suzaku} counterpart). However, the source might be just too blurred in the \emph{ASCA} images. 

Neither the \emph{Suzaku} compact source nor the large-scale extended emission were detected in the 2013 \emph{CXO} observation. To estimate an upper limit on the count rate of the compact source or large-scale emission, we searched for the highest numbers of counts in multiple conservatively chosen apertures with $r=3\arcsec$ within $20\arcsec$ of the \emph{Suzaku} source position (18:11:51, $-$19:00:54). We found $n_{\rm max}=9$ (source and background) counts, which correspond to a 99\% confidence upper limit of $n_{ul}=18.8$\,counts (see Table\,1 in \citealt{Gehrels1986}). Considering further the $n_{\rm BG}=675$\,counts in a $r=50\arcsec$ background aperture, we obtain a 99\% confidence upper limit on the source count rate, $R_{s,ul}= 3 \times 10^{-4}$\,counts\,s$^{-1}$. If the same spectral parameters and the same X-ray source flux are assumed as obtained with \emph{Suzaku}, we would expect  $R_{s,exp}= 73\pm3 \times 10^{-4}$\,counts\,s$^{-1}$ in the 2013 \emph{CXO} observation. Thus, the ``compact'' source is a transient source with a flux variation of at least a factor of 24.

In the region of the \emph{Suzaku} extended emission \texttt{wavdetect} found several ($\approx 40$) faint sources in the 2013 \emph{CXO} observation. Applying the CIAO task \texttt{srcflux}, we estimate that the combined X-ray source flux of the faint \emph{CXO} sources is $1.0^{+0.4}_{-0.2} \times 10^{-12}$\,erg\,cm$^{-2}$\,s$^{-1}$ in the energy range 0.5--10\,keV. Fitting the \emph{Suzaku} extended emission spectrum with a PL, we obtained $n_{H, 22}=1.4^{+0.6}_{-0.5}$, $\Gamma=1.7\pm 0.3$, $\chi_\nu^2=1.3$ for $\nu=76$ dof, and derive an absorbed flux of $1.0^{+0.07}_{-0.13} \times 10^{-12}$\,erg\,cm$^{-2}$\,s$^{-1}$  in the energy range 0.5--10~keV. Thus, we can explain the \emph{Suzaku} extended emission entirely by the fainter point sources resolved with \emph{CXO}.

\subsection{Radio}

1.4~GHz images of the TeV ``extension'' (Figure~5) reveal the diffuse shell with the diameter of $\sim6\arcmin$. In the Green's catalog \citep{Green2009} it is categorized as SNR G11.4$-$0.1 which is likely to be at a distance of 6-14 kpc  \citep{Brogan2004}. We searched for diffuse emission in the \emph{CXO} (ObsID 3512) and \emph{Suzaku} (ObsID 504078010) images, but the SNR was not detected in X-rays. The bright radio point source at the southern edge of G11.4$-$0.1 has a flat radio spectrum ($\alpha\sim0.1$), and, since it does not have an infrared counterpart, it is most likely an unrelated extragalactic source \citep{Brogan2004}. Because of its brightness in radio, it is unlikely that the source is an undetected pulsar. A few other faint radio point sources are seen in the 1.4~GHz images, but they have no counterparts at other wavelengths either.

\begin{figure}
\includegraphics[scale=0.305,trim=0 0 0 0]{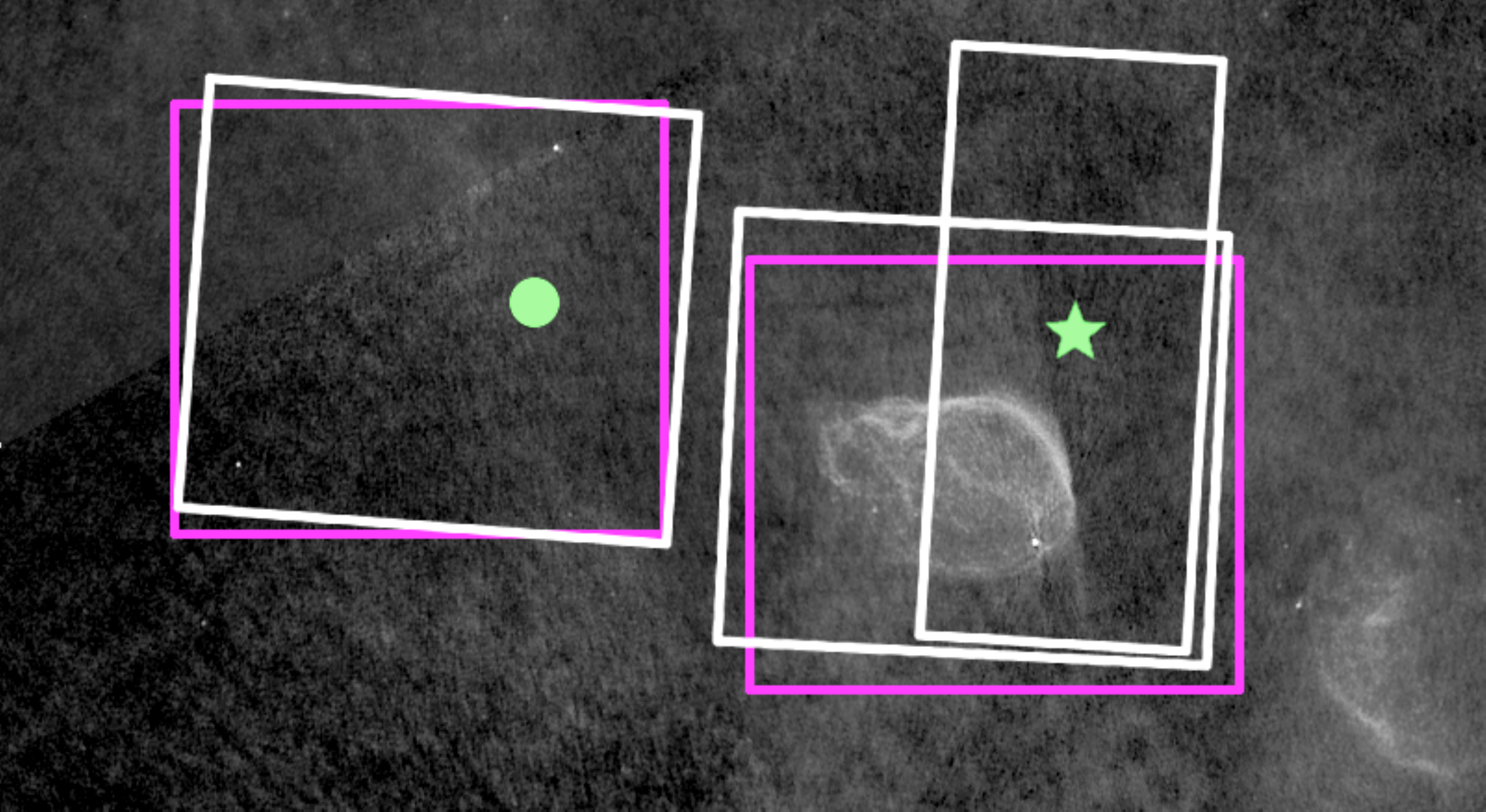}
\caption{1.4 GHz VLA image of the TeV NE extension. The purple and white shapes show the coverage of the \emph{CXO} (ACIS-I and HRC-S) and \emph{Suzaku} observations, respectively (same as those shown in the Figure~1 left panel).}
\end{figure}

\subsection{Fermi Data}

We do not find firm evidence of GeV $\gamma$-ray emission from 2FGL J1811.1-1905c in the {\sl Fermi} LAT data, and place an upper limit to the $\gamma$-ray flux, $F_\gamma=(1.8\pm0.2)\times10^{-10}$~erg~cm$^{-2}$~s$^{-1}$ in the 0.3--200~GeV band. We used the PL fit with $\Gamma=2.7\pm0.1$ to estimate this limit. The 2FGL\footnote{\url{http://fermi.gsfc.nasa.gov/ssc/data/access/lat/2yr_catalog/}} catalog lists only two ``confused'' sources\footnote{A ``c'' following the 2FGL name indicates that the source is found in a region with bright and/or possibly incorrectly modeled diffuse emission.} in the region. These sources are shown on the right panel of Figure~1 with 95\% error ellipses. We find marginally significant (3.1$\sigma$--3.7$\sigma$, depending on the background choice) excess in the 15-300 GeV range for the region shown by the dashed circle in Figure~1. The excess is, however, noticeably offset from the previously reported 2FGL 1811.1-1905c source position.

\section{Discussion}

{\bf XTE~J1810--189:} XTE~J1810--189 has been observed on a number of occasions by multiple observatories such as \emph{CXO}, \emph{Suzaku}, \emph{Swift}, and \emph{RXTE}. The LMXB shows both quiescent and outburst periods (see Figure~3). The compact object was identified as a Type I X-ray burster \citep{Markwardt2008} based on the properties of its 2008 outburst.

The typical luminosity of LMXBs in quiescent state (qLMXBs) is $L_X\approx10^{31}-10^{33}$~erg~s$^{-1}$ (0.5--10~keV; \citealt{Heinke2003}). LMXBs containing NSs can be confidently identified if they experience bright type-I X-ray bursts caused by unstable thermonuclear burning of the accreted matter on the NS surface. These transiently accreting NSs  usually show a soft, blackbody-like X-ray spectral component, and/or a harder X-ray component generally fit by a PL with photon index $\Gamma=1-2$. A quiescent state of LMXB could explain the 2003 \emph{CXO} non-detection. The flux variability of XTE J1810$-$189, by a factor of $>10^6$, is among the highest recorded but still  is consistent with those seen in other LMXBs \citep{2009A&A...495..547D}.

XTE~1810--189 is situated in the Galactic disk, and with A$_{\rm V}\ge11.5$ it is not surprising that it is not detected in the optical (the 2004 CTIO observations). However, NIR variability was detected \citep{Torres2008}. 

{\bf Suzaku~J1811.} Based on its X-ray spectral properties, the source could be a transient X-ray binary pulsar or a magnetic cataclysmic variable (CV). X-ray binary pulsar spectra at the \emph{Suzaku} XIS energy range have been described as including a PL component with photon index $\Gamma\sim 1$, a soft BB component with a temperature of  $kT \approx 90-300$\,eV, and an iron emission line at energies $6.4-6.7$\,keV (e.g., \citealt{Hickox2004}). CVs show lower BB temperatures than those seen in most X-ray binary pulsar spectra, and a soft thermal excess with $kT \approx 20-40$\,eV is particularly found for magnetic CVs (e.g., polars; \citealt{Warner2003}). A prominent emission line at 6.7\,keV, presumably from helium-like ionised Fe in hot plasma can also be seen in almost all magnetic CVs \citep{Singh2013}. The temperature of J1811's BB component, $kT=60\pm 20$~eV, is inbetween the ranges of CVs and X-ray binary pulsars.

The range of J1811's effective thermal emitting radius, $R_{BB}= 540 D_{10}$\,km to $2 \times 10^{5} D_{10}$\,km, excludes neither magnetic CV nor X-ray binary pulsar interpretation. In polars, the soft emission is likely produced in hot spots on the white dwarf through absorption and reprocessing of the hard X-ray emission from the accretion column (e.g., \citealt{Warner2003}). In X-ray binary pulsars with X-ray luminosities similarly low as the one of J1811 (see below), \citet{Hickox2004} suggested that the soft thermal excess is produced by diffuse gas in the system\footnote{A PL$+$mekal$+$ iron line also fits the {\emph{Suzaku}} data with reduced\,$\chi^2=0.98$ for 71 dof.}. In both cases, the sizes of the emission regions would be consistent with the poorly constrained thermal emission area of J1811.

The 0.3--10 keV absorbed luminosity of J1811 in the \emph{Suzaku} data is in the range $2 \times 10^{32}$\,erg\,s$^{-1}$ to $4 \times 10^{33}$\,erg\,s$^{-1}$ if the source distance is 2\,kpc to 10\,kpc. Such a low luminosity favors a CV rather than X-ray pulsar classification since for the latter luminosities typically exceed $10^{34}$\,erg\,s$^{-1}$ (e.g., \citealt{Wijnands2006}). However, {\emph{Suzaku}} might not have caught J1811 at its peak luminosity. Therefore, we cannot firmly rule out the possibility of the X-ray binary pulsar scenario.

\emph{Suzaku}~J1811 was not seen in the 2013 \emph{CXO} observation. The closest \emph{CXO} source is $\sim 58\arcsec$ away from the center of the compact \emph{Suzaku}~J1811 source. This X-ray source is close to source \#5 in Figure~2. It is not included in Table~2 because its detection significance and its net count number are lower than our chosen thresholds (see beginning of Section 3). Given the positional uncertainty of \emph{Suzaku} of $\sim20\arcsec$ (see \emph{Suzaku} ABC Guide\footnote{\url{http://heasarc.gsfc.nasa.gov/docs/suzaku/analysis/abc/abc.html}}), it is unlikely that this X-ray source is the \emph{CXO} counterpart. The found X-ray flux variation of (at least) factor 24 (Section 3.2) is not unusual for CVs or for accreting pulsars (e.g., \citealt{Lin2012}; their Figure 4c).

We cannot realiably identify the MW counterpart of J1811 because of the positional uncertainty of \emph{Suzaku} and the non-detection by \emph{Chandra}. One of several NIR sources, and, in particular, a highly reddened \emph{Spitzer} source might be the counterpart.  

It remains unclear whether  \emph{Suzaku}~J1811 could contribute to the TeV emission in the NE part of HESS J1809$-$193. Some high-mass X-ray binaries (HMXBs) show $\gamma$-ray emission up to 10~TeV which is usually interpreted as coming from colliding wind shocks (e.g., \citealt{Dubus,Aharonian2006B,2014AN....335..301K}). However, the X-ray spectra of these HMXBs do not show significant thermal components or iron lines (e.g., \citealt{Aliu2014,Chernyakova2006,Martocchia2005}) -- in contrast to the spectrum of J1811. To our knowledge, there is no known case of a CV with TeV emission\footnote{see, e.g., the TeV source catalog by Wakely \& Horan,\protect\\  \url{http://tevcat.uchicago.edu}}. If the putative $\gamma$-ray emission NE of HESS J1809-193 is confirmed, J1811 could be a new type of TeV binary.

In summary, the X-ray properties of \emph{Suzaku}~J1811 are consistent with those of X-ray binary pulsars or magnetic CVs. However, considering the low X-ray luminosity, a magnetic CV appears to be the most likely counterpart of J1811.

From the comparison of our flux estimate for the point sources in the \emph{CXO} image with the flux estimate for the extended emission seen in the \emph{Suzaku} image, we conclude that the latter was merely a low spatial resolution effect, i.e., there is no statistically significant extended X-ray emission around J1811.

{\bf \emph{CXO} X-ray point sources:} We analyzed the MW properties (X-ray, optical, NIR and IR photometry  from available surveys) of the 16 sources detected in the \emph{CXO}/ACIS-I images, and classified them using an automative algorithm \citep{Brehm}. The algorithm constructs a decision tree from a training dataset consisting of known objects of 9 classes -- AGNs, LMXBs, HMXBs, main sequence stars, Wolf-Rayet stars, cataclysmic variables (CVs), isolated NS, young stellar objects (YSOs), and non-accreting binary pulsars. The decision tree is then applied to the sample of unknown X-ray sources. Our classification of the 16 X-ray sources produced only six classifications with confidence $>70\%$ (Table~2), including 4 stars, 1 LMXB, and 1 YSO. After careful examination, none of the 16 X-ray sources appears to be a convincing candidate for the source of the TeV emission.

{\bf Relation to HESS J1809--193:} The spatial morphology and the extent of the putative TeV emission NE  of HESS J1809$-$193 are rather uncertain and may include XTE J1810$-$189 position. However, XTE J1810$-$189 is an ordinary Type I X-ray burster, and so far such objects have not been found to produce TeV $\gamma$-rays. SNR G11.4$-$0.1 is also too offset from the TeV source, and is relatively small in size. Suzaku J1811 is close to the center of the faint TeV extension. The X-ray source appears to be transient, as no counterpart is found in the later  ACIS-I image of Suzaku J1811 field. The extended X-ray appearance of this source in  \emph{ Suzaku} XIS can be explained by the multiple point sources (found in the CXO images) smeared by the wide PSF of  \emph{ Suzaku} XRT. Therefore, no truly X-ray diffuse emission is detected in the region. We suspect that Suzaku J1811 is a magnetic CV, or less likely an accreting pulsar. Therefore, it is not expected to produce TeV $\gamma$-rays based on our current knowledge of these objects. Our MW classification of the 16  \emph{ CXO} sources also did not yield a promising candidate for the putative TeV emission NE of HESS J1809-193.

The flux limit derived from the \emph{Fermi} data is consistent with the simple extrapolation of the observed HESS J1809--193 spectrum (with $\Gamma=2.2\pm0.1$; Aharonian et al.\ 2007) to the {\sl Fermi} band if the TeV emission from the NE constitutes $\sim1$\% of the HESS J1809--193 flux (an approximation consistent with Figure~1 from Aharonian et al. 2007). The TeV NE extension would have a $\gamma$-ray flux in the \emph{Fermi} band (0.2-300 GeV) of $F_\gamma=1.4\times10^{-12}$~erg~cm$^{-2}$~s$^{-1}$. While the simple PL extrapolation is in agreement with the estimated \emph{Fermi} flux limit, it is also possible that the IC spectrum is affected by cooling, which would happen if the electron Lorentz factors exceed $\gamma\sim8\times10^{7}[1+0.144(B/1~{\rm \mu G})^2]^{-1}(\tau/50~{\rm kyrs})^{-1}$ for the continuous electron injection within the source with the age $\tau$, in the magnetic field $B$ and the radiation field of CMB (de Jager \& Djannati-Atai 2009). In this case, the Fermi spectrum produced by ICS on CMB photons can be a harder PL with the slope photon index $\Gamma\sim1.7$ and a substantially smaller GeV flux. A deeper HESS exposure needs to be analyzed before any further meaningful conclusions can be made.

\section{Conclusions}

We studied multiwavelength data for the region located in the outskirts of the TeV source HESS J1809--193.
Several interesting  X-ray sources have been detected but none of them is a promising candidate for the putative TeV emission NE of the HESS J1809--193 bright part. The remote radio SNR~G11.4$-$0.1 is not detected in X-rays and is unlikely to be a detectable TeV source. Using $\sim6$ years of \emph{Fermi} LAT data we did not detect the 2FGL 1811.1-1905c at its catalogued position although we found tentative evidence of hard GeV emission at somewhat different location (farther from HESS J1809--193 center), which has not been covered by X-ray observations yet. Deeper $\gamma$-ray and X-ray observations of this interesting region  may be warranted.

\acknowledgements 

Support for this work was provided by the National Aeronautics and Space Administration through Chandra Award Numbers GO3-14049X issued by the Chandra X-ray Observatory Center, which is operated by the Smithsonian Astrophysical Observatory for and on behalf of the National Aeronautics Space Administration under contract NAS8-03060. This research was partially supported by NASA ADAP grants NNX10AH82G, and NNX09AC81G. This publication makes use of the extinction map query page, hosted by the Centre for Astrophysics and Planetary Science at the University of Kent.

%%%%%%%%%%%%%%%%%%%%

\end{document}